\documentclass[twocolumn,showpacs,amsmath,amssymb]{revtex4}
\usepackage{epsfig}

\begin{document}

\title{Unwrapping of DNA-protein complexes under external stretching}
\author{Takahiro Sakaue(1) and  Hartmut L\"owen(2)}
\affiliation{1)Department of Physics, Graduate School of Science, Kyoto University, Kyoto 606-8504, Japan,
2) Institut f{\"u}r Theoretische Physik II, Heinrich-Heine-Universit{\"a}t D{\"u}sseldorf, 
Universit{\"a}tsstra{\ss}e 1, D-40225 D{\"u}sseldorf, Germany}

\begin{abstract} 
A DNA-protein complex modelled by a semiflexible chain and an  attractive spherical core is studied in the situation when an external stretching force is acting on one end monomer of the chain while the other end monomer is kept fixed in space. 
Without stretching force, the chain is wrapped around the core. 
By applying an external stretching force, unwrapping of the complex is induced.
We study the statics and the dynamics of the unwrapping process by computer simulation and simple phenomenological theory. We find two different scenarios depending on the chain stiffness:
For a flexible chain, the extension of the complex scales linearly with the external force applied.
The sphere-chain complex is disordered, i.e.\ there is no clear winding of the chain around the sphere.
For a stiff chain, on the other hand, the complex structure is ordered, which is reminiscent to nucleosome.
There is a clear winding number and the unwrapping process under external stretching is discontinuous with jumps of the distance-force curve. 
This is associated to discrete unwinding processes of the complex. 
Our predictions are of relevance for experiments, which measure force-extension curves of DNA-protein complexes, such as nucleosome, using optical tweezers.

\end{abstract}


\maketitle
\section{Introduction}
Complexations between chain-like molecules and spherical host particles are frequent in nature.
Key examples can be found in living organism, where various DNA-protein complexes play important roles in fundamental life processes \cite{DNA-protein}. 
This complex formation is also important for potential applications in gene delivery, where negatively charged DNA is complexed with various cationic substances to be an efficient gene carrier.
One of the prominent examples is the nucleosome, which is constituted by cationic proteins, called histone octamer, and DNA wrapping around them.
The formation of the nucleosome is the first step for a dramatic compaction of long DNA chains from  the order of $cm$ down to a nucleus size of $\sim$ $\mu$ m \cite{chromatin_book2,chromatin1,chromatin_book1}.
Since the tight wrapping of DNA around the histone should limit the accessibility of transcriptional factors, the nucleosome structure in genetically active states is expected to be loosened or, at least, partially unwrapped.
Therefore, the stability and dynamical properties of the nucleosome are crucial factors for the gene activity in eukaryotic cells \cite{Workmann_Annu.Rev.Biochem,Schiessel_JPCM_review}.

Theoretically, simple systems with a spherical or cylindrical particle and a polymer chain have been adopted to capture the essential features of nucleosomes \cite{Marky_BioP,Wallin_Langmuir,Zhang_JPC,Mateescu_EPL,Netz_MacroM,Kunze_PRL,Kunze_PRE,Schiessel_EPL,Sakaue_PRL,Jonsson_JCP,Akinchina_MacroM,Schiessel_JPCM_review}.
The sphere-chain complexation behaviour is controlled by several factors, such as the nature of interaction between them, the sphere size, and the chain stiffness.
It is an interesting question to study the response of the complex with respect to an external force acting on the chain \cite{Kunze_PRE,Kulic_cm}. A dramatic change of the complex structure is expected for large external forces.
In fact, it is known that the wrapping behavior of nucleosomes in the cell is influenced by the tension generated by molecular motors \cite{Cook_Science,Leuba_PNAS}.

In the present article, we study the complexation behavior of a simple 
DNA-protein model system by means of computer simulations and a phenomenological theory.
We do not restrict ourselves to the nucleosome-like structure only, but aim to get the full picture of properties inherent to the sphere-chain complex under the tension.
In particular, we investigate the effect of chain stiffness on the structure of a sphere-chain complex, and its consequent characteristics in response to external tension acting on the end monomers of the chain.
We find two different scenarios, which depend crucially on the chain stiffness:
For a flexible chain, the extension of the complex scales linearly with the external force applied.
The sphere-chain complex is disordered, i.e.\ there is no clear winding of the chain around the sphere.
In the opposite limit of a stiff chain, however, the complex takes an nucleosome-like ordered structure and there is a clear winding number.
The unwrapping process under external tension is discontinuous exhibiting jumps in the distance-force curve. 
This can be traced back to discrete unwinding processes of the complex. Our predictions are of relevance for experiments, which measure force-extension curves of DNA-protein complexes, such as nucleosome, using optical tweezers \cite{Leuba_PNAS,Cui_PNAS,Bennink_NatureStBio,Brower-Toland_PNAS}.

Our paper is organized as follows: In chapter II, we introduce a monomer-resolved model of DNA-protein complexes which is studied by simulation. Simulation results are presented in chapter III. 
A further discussion based on a simple theory is contained in chapter IV. Finally
we conclude in chapter V.

\section{Simulation Model}
The model we adopted is almost the same as that studied previously in Ref.\  \cite{Sakaue_PRL}.
First, a polymer chain is represented by  $N=50$ spherical monomers at positions $\mbox{\boldmath$r$}_i$.
In what follows, we rescale all lengths and energies by the size of monomer diameter $b \equiv 1$ and thermal energy $k_B T \equiv 1$, respectively.
Neighbouring monomers are connected via harmonic bonds; the bond energy between neighbours reads as  $U_{bond}=\frac{\kappa_{bond}}{2}(|\mbox{\boldmath$r$}_i - \mbox{\boldmath$r$}_{i+1}| -1)^2$.
We set the spring constant $\kappa_{bond}=400$, which gives almost a constant bond length $\sim 1$.
The chain stiffness is implemented by the bending potential of the form $U_{bend}=\kappa \left\{ 1-(\mbox{\boldmath$r$}_{i-1}-\mbox{\boldmath$r$}_{i})(\mbox{\boldmath$r$}_{i}-\mbox{\boldmath$r$}_{i+1}) \right\}$, which favors stretched configurations.
The bending modulus $\kappa$ is connected with the chain persistence length $\lambda_p$ via $\lambda_p \simeq \kappa-0.5$ (for $\kappa > 2$) \cite{Sakaue_PRL}.
The excluded volume of monomers is modelled by the repulsive part of a Morse potential, $U_{M,rep}(r_{i,j})=\epsilon_m \exp{\left\{ -\alpha_m(r_{i,j}-1) \right\}}$, where $\epsilon_m = 1.0$, $\alpha_m =24$ and $r_{i,j}$ is the distance between monomers.

Next, we introduce a spherical particle.
The interaction between each monomer and the spherical particle is modelled through the full Morse potential, $U_{M}=\epsilon[\exp{\{-2 \alpha (r_i -\sigma) \} } -2 \exp{\{-\alpha (r_i-\sigma) \}}]$, where $r_i$ denotes the distance between $i$ th monomer and the spherical core.
We set $\alpha=6$ and $\sigma=1.9$ (corresponding to a sphere diameter of about 2.6), so that the volume ratio between the spherical core and polymer chain mimics a real nucleosome\cite{Sakaue_PRL}.
The strength of attraction is fixed to be $\epsilon = 8$.
It should be noticed that such a model with only short range interactions would be readily accepted for the charged system, too, in the salty environment (including the physiological condition), where a long range electrostatic interactions become local due to the screening.

\begin{figure}[htb]
\epsfig{figure=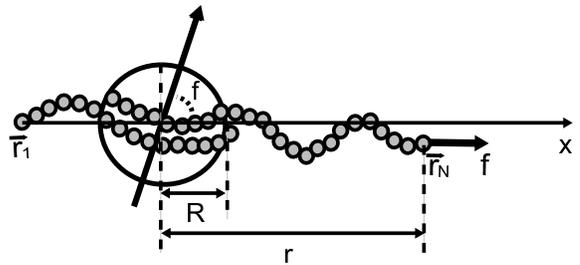,width=7.5cm}
\caption{Schematic sketch of the complex under external stretching. The first end monomer is fixed at $\mbox{\boldmath$r$}_1$, the protein sphere with radius $R$ is fixed at the origin and the other end monomer is pulled by the force $f$ along the $x$ axis.
The chain extension $r$, and the tilt angle $\phi$, which can be defined for the ordered complex, are also shown.}
\label{schematic}
\end{figure}

Even though the tightly wrapped complex is formed, the spherical core slides along the chain and prefers positioning at the chain end\cite{Sakaue_PRL}.
We would like to put the complex now under an external tension.
In order to get a clearcut configuration we fix simultaneously one end monomer {\it and} the centre of the sphere:
the spherical core is fixed at the origin, and the end monomer at $\mbox{\boldmath$r$}_1 = (-8,0,0)$.
Then, stretching force is applied to the other end monomer with position  $\mbox{\boldmath$r$}_N$ along the $x$ axis (see Fig. \ref{schematic}).
We realize this by either applying a constant force $f$ or by stretching the monomer at a constant velocity $v$.
In the former case, we obtain the average extension in $x$ direction
$ r  \equiv \langle r_{N,x} \rangle$, while in the latter, we monitor the resistant force during the possibly nonequilibrium, unwrapping process.

The monomers obey the stochastic dynamics described by an underdamped Langevin equation.
\begin{eqnarray}
m \frac{d^2 \mbox{\boldmath$r$}_i}{dt^2} = -\gamma \frac{d \mbox{\boldmath$r$}_i}{dt} +\mbox{\boldmath$R$}_i (t) -\frac{\partial U}{\partial \mbox{\boldmath$r$}_i } .\ 
\end{eqnarray}
The role of the solvent is incorporated by the friction ($\gamma$ is the friction constant) and the random kicks acting on the particles.
This random force is represented by a Gaussian delta-correlated noise whose variance is related to the friction constant through the fluctuation-dissipation theorem: $\langle \mbox{\boldmath$R$}_i(t) \mbox{\boldmath$R$}_j (t') \rangle = 6 \gamma \delta_{ij} \delta(t-t')$.
$m$ is the monomer mass and the total internal energy $U$ splits naturally into $U = U_{bond}+U_{bend}+U_{M, rep}+U_{M}$. 
We integrate the discretized equation of motion using a leap-frog algorithm with time step of $\Delta t = 0.0025 \tau$, where $\tau (= \gamma b^2/k_BT)$ is a typical time scale of monomer diffusion. 
In the following, the time is represented with a rescaled unit with $\tau$. We remark that hydrodynamic interactions as induced by the solvent flow are neglected in our approach, but these are irrelevant if equilibrium quantities are calculated.

\section{Results}
\begin{figure}[htb]
\epsfig{figure=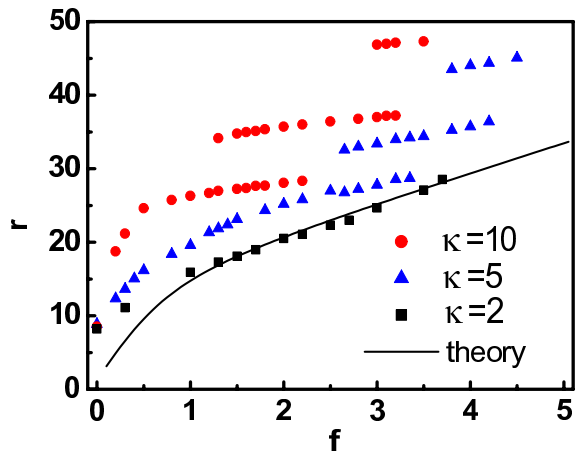,width=7.5cm}
\caption{(Dot) Force-extension relation of the sphere-chain complex obtained by computer simulations for different chain stiffnesses $\kappa =2$, $5$ and $10$. In case of a bimodal distribution of $r$, we plot both peak positions. (Line) Theoretical force-extension relation of the disordered sphere-chain complex. The parameters are estimated from the corresponding system in the simulation with $\kappa = 2$: $a_1 =-7$, $a_2= 0.12$, $L_0 = 44$, $\lambda_p = 1.5$. }
\label{f-r}
\end{figure}

We first study the response of the complex to the applied constant tension.
We carried out sufficiently long simulations, typically $\sim 4 \times 10^8$ time steps, to extract equilibrium properties of the ensemble.
In Fig. \ref{f-r}, we show the force-extension relation for the cases with different three chain stiffnesses, $\kappa = 2$, $5$ and $10$.
In all cases, when the applied tension is weak, the extension increases rather rapidly with the tension.
However, the response under moderate or high tension crucially depends on the chain stiffness.
The extension of the complex with a flexible chain ($\kappa = 2$) gradually increases with the applied tension with almost {\it linear} dependency.
On the other hand, if the chain is stiff enough ($\kappa = 10$), the extension is almost constant, in other words, the susceptibility is very low until the abrupt discrete change occurs around some critical tension.
It can be shown that this distinction comes from the structural property of the complex.

\begin{figure}[htb]
\epsfig{figure=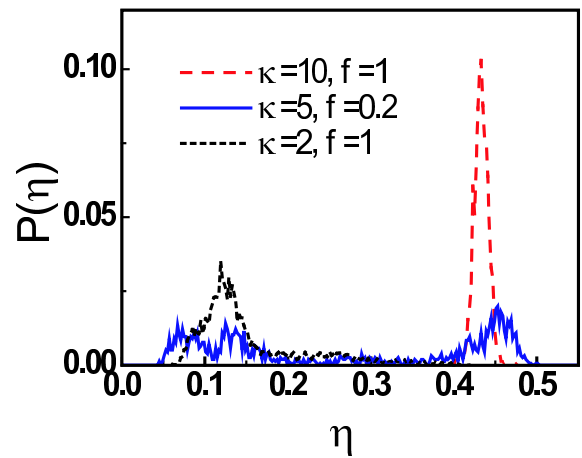,width=7.5cm}
\caption{Probability distribution of the order parameter $\eta$ for complexes made from chains with different stiffnesses $\kappa =2$, $5$ and $10$ under a stretching force $f$.}
\label{eta}
\end{figure}

\begin{figure}[htb]
\epsfig{figure=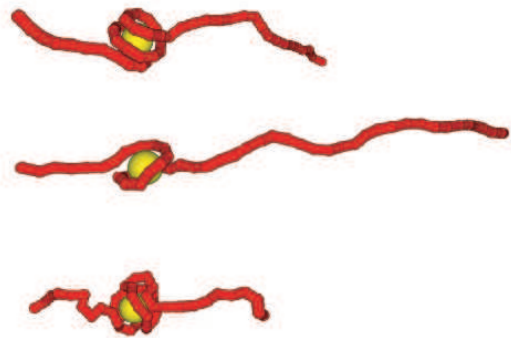,width=7.5cm}
\caption{Typical simulation snapshots of the sphere-chain complex. Top: an ordered wrapping with wrapping number $w=2$ for $\kappa = 10$, $f=1$. Middle: an ordered wrapping with $w=1$ for $\kappa=10$, $f=2.5$. Bottom: a disordered state for $\kappa=2$, $f=1$.}
\label{snapshot}
\end{figure}

To see this, we introduce the following parameter \cite{Sakaue_PRL},
\begin{eqnarray}
\eta \equiv\frac{|\sum\limits_{<i>} \mbox{\boldmath$r$}_{i,i+1} \times  \mbox{\boldmath$r$}_{i+1,i+2} |}{N_{p}} ,
\label{orderparameter}
\end{eqnarray}
where the summation in $\eta$ is taken only over the monomers in the vicinity of the particle's  surface  $(r_{i}<2.9)$, and $N_{p}$ is the number of such monomers.
This quantity measures a degree of ``wrapping ordering'' of the complex.
If the chain wraps orderly and traces the helical path on the sphere surface, $\eta$ takes a high value of $\eta \gtrsim 0.4$.
Otherwise, the complex is regarded as a random adsorption of the chain on the sphere, which results in the small value of $\eta \lesssim 0.2$.
Figure \ref{eta} shows the probability distribution of $\eta$ for complexes of different chain stiffnesses.
The ordered and disordered complexes are formed from the stiff ($\kappa=10$) and flexible ($\kappa=2$) chains, respectively, and the bimodal distribution between these two states is observed in the case of intermediate chain stiffness ($\kappa=5$).
Figure \ref{snapshot} shows typical simulation snapshots of the complex corresponding to ordered and disordered states.
Note that $\eta$ for the complex with one-turn (wrapping number $w=1$) is about $\eta \simeq 0.33$, which is slightly smaller than that for the complex with $w=2$ ($\eta \gtrsim 0.4$).

\begin{figure}[htb]
\epsfig{figure=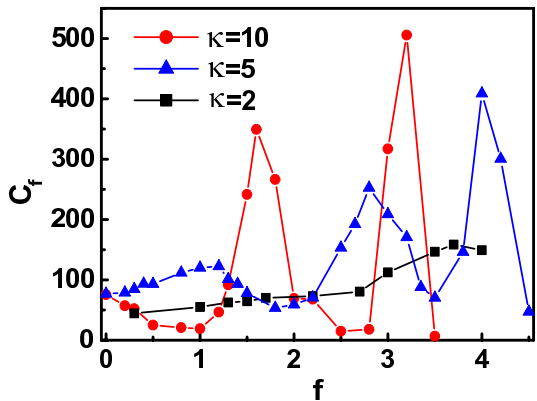,width=7cm}
\caption{Specific heat as a function of the stretching force for different chain stiffnesses, $\kappa=2$, $5$ and $10$.}
\label{SH}
\end{figure}

Figure \ref{SH} shows the specific heat $C_f$ for various chain stiffnesses as a function of the  applied stretching force $f$.
In the constant force ensemble, it is the temperature derivative of the enthalpy, which is calculated as $C_f = \langle U ^2 \rangle -\langle U \rangle ^2 - f (\langle r_{N,x} U \rangle -  r \langle U \rangle)$.
For a flexible chain ($\kappa =2$), $C_f$ just slightly increases with $f$ in a monotonic way, while for stiffer chains, it shows sharp peaks around $f=1.6,\ 3.2$ for $\kappa=10$, and $f=2.8,\ 4$ for $\kappa=5$.
These peaks exactly correspond to the points, where the force-extension relation reveals abrupt jumps, and indicate the transition from the wrapping number $w=2$ to $1$, and $w=1$ to $0$.
In the remanent values of an applied tension $f$, $C_f$ is very small and $r$ is almost constant.
In contrast to the lax response of the disordered complex from a flexible chain, the response of the ordered complex is characterized by a ``switching behavior'': 
it is very stable against tension, but reveals the sudden switching to a different state by an unwrapping transition at some critical tension.
This trend becomes more evident for the complex with the stiffer chain.

\begin{figure}[htb]
\epsfig{figure=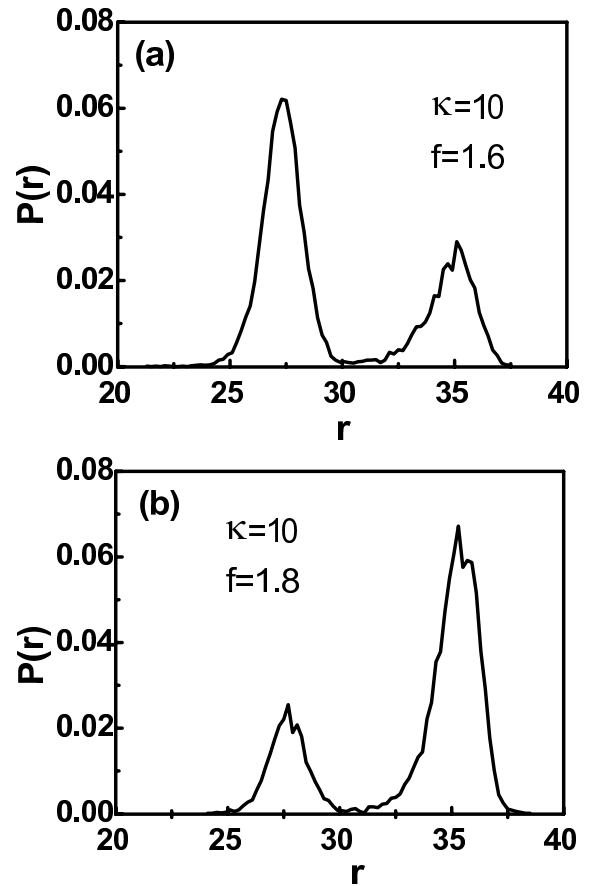,width=7.5cm}
\caption{Probability distribution of the chain extension $r$ (a) below (f=1.6) and (b) above (f=1.8) the critical stretching force to induce the unwrapping transition from $w=2$ to $w=1$ for the chain stiffness of $\kappa=10$.}
\label{r-dist}
\end{figure}

In Fig. \ref{r-dist}, we show the probability distribution of the chain extension $r$ around the critical force of the transition from $w=2$ to $w=1$ for $\kappa = 10$.
Clear bimodal distributions are seen, where the relative probability changes with the increase in $f$, whereas, the peak positions remain unchanged. 
The free energy barrier between these two states is deduced to be $\Delta G \simeq 4 \sim 5$. 

For the chain with $\kappa=5$, an additional, somewhat broad peak in $C_f$ is seen around $f \simeq 1.2$.
The analysis within the distribution of $\eta$ indicates that this is a signal for a tension induced ordering transition.
As Fig. \ref{eta} shows, the complex with intermediate stiffness ($\kappa=5$) reveals the bimodal distribution of $\eta$ under the weak tension.
This equilibrium between disordered and ordered states shifts to ordered one with the increase in $f$.
At $f \simeq 1.5$, the distribution of $\eta$ becomes similar to that of ordered complex.
Under the stronger tension, the complex behaviors are almost identical to those of ordered one.

\section{Discussions}
\subsection{Force-extension relation in a disordered complex}
\label{sec_disorder}
Now, we present a phenomenological theory to describe the properties of the sphere-chain complex in a disordered state under the stretching force.
We consider a polymer chain (length $L_0$, thickness $b=1$), one part of which (its length is $L$) is adsorbed on the spherical core of radius $R$.

The remaining part of the chain (length $L' = L_0-L$) is free and unbound by the core particle. It takes a stretched conformation due to the stretching tension applied to the chain end, whose end-to-end distance (distance from the core surface to the stretched chain end) is denoted as $r'$.

In the disordered state, there are substantial freedoms for the adsorbed segment distribution on the core surface.
To describe this situation, we adopt a Flory-type mean-field argument.
A conditional free energy of the disordered state characterized by an adsorbed length $L$, under the action of a stretching force $f$ is written as
\begin{eqnarray}
G_d(L) = F_d(L) +F_u(L'),
\label{G_disorder}
\end{eqnarray}
where $F_d$ and $F_u$ represent the free energy of the disordered segments on the core surface and that of unbound segments under tension.

$F_d(L)$ is composed of the following contributions.
\begin{eqnarray}
F_d(L)= F_{ad} +F_{conf}+F_{ss}.
\label{F_disorder}
\end{eqnarray}
The first term is the energetic gain due to the segment adsorption on the core: $F_{ad}=\epsilon L$, where $\epsilon <0 $ is the adsorption energy per length.
The second term is the conformational entropy of the chain on the core, which is evaluated from the random walk on the core (proportional to the segment number $L/\lambda_p$) plus its deviation from the ideal chain statistics: $F_{conf} \simeq -L/\lambda_p +R^2/L\lambda_p +L\lambda_p/R^2$\cite{RedBook}.
The last term in eq. (\ref{F_disorder}) describes the segment-segment interaction, which can be taken into account through the virial expansion up to second order: $F_{ss} \simeq B n^2 R^2$, where $B$ is the second virial coefficient $B \sim b \lambda_p$, and $n$ stands for the segment density on the core $n \sim L/R^2\lambda_p$.

We describe the unbound chain under tension by freely jointed chain with the segment length of $2 \lambda_p$, which writes the free energy $F_u$ as \cite{RedBook}
\begin{eqnarray}
F_u (L') = - \frac{L'}{2 \lambda_p} \ln{\left( \frac{4 \pi \sinh{y}}{y} \right)},
\label{F_unbound}
\end{eqnarray}
where $y \equiv 2 \lambda_p f$.
The end-to-end distance $r'$ is associated with the contour length of unbound part $L'$ as
\begin{eqnarray}
r' = L' \left( \coth{y} -\frac{1}{y} \right) .
\label{r-f_freely-jointed}
\end{eqnarray}

By putting eq. (\ref{F_disorder}) and eq. (\ref{F_unbound})  in order, we obtain
\begin{eqnarray}
G_d(L) = a_1 L + a_2 L^2 +a_3 L^{-1} \nonumber \\
-\frac{L_0-L}{2 \lambda_p}\ln{\left(\frac{4 \pi \sinh{y}}{y} \right)},
\label{G_disorder2}
\end{eqnarray}
where $a_1 \simeq \epsilon + \lambda_p/R^2$, $a_2 \simeq b/\lambda_p R^2$, and $a_3 \simeq R^2/\lambda_p$.
The third term is shown to be always negligible compared to the second term in the case of our interest, thus we can safely neglect it.
And note that $L=L_0 -L'$ is a function of $r'$ (see eq. (\ref{r-f_freely-jointed})).

The most probable length for the adsorbed chain part is determined through the minimization of the free energy, $\partial G_d / \partial r' = (\partial G_d / \partial L)( \partial L / \partial r') = 0$ (This is equivalent to the equality of the chemical potentials of adsorbed and unbound segments).
\begin{eqnarray}
\langle L \rangle =-\frac{a_1}{2a_2} -\frac{1}{4 a_2 \lambda_p} \ln{\left( \frac{4 \pi \sinh{y}}{y}\right)}.
\label{L_disorder}
\end{eqnarray}
To compare with the result from simulations, it is convenient to rewrite this relation in terms of $r$ using $r=r'+R$.
\begin{eqnarray}
 r  = \left[ L_0 + \frac{a_1}{2a_2}+ \frac{1}{4 a_2 \lambda_p}\ln{\left( \frac{4 \pi \sinh{y}}{y}\right)}\right] \nonumber \\
\times \left[ \coth{y} -\frac{1}{y} \right] +R, 
\label{r-L_theory}
\end{eqnarray}
This function is depicted in Fig. \ref{f-r}.
All of the coefficients in eq. (\ref{r-L_theory}) can be estimated once the system is specified.
Therefore there is, in principle, no space for adjustable parameters.
The theoretical prediction shows a good agreement with the simulation result, especially for large forces.

Let us consider the two limiting cases, (i) small force regime ($y \ll 1$) and (ii)large force regime ($y \gg 1$).
By taking asymptotic limits in eq. (\ref{r-L_theory}), we obtain
\begin{eqnarray}
r  = 
\begin{cases}
b_1 + b_2 f & (y \ll 1), \\ 
\label{r-L_theory_largeF}
b_3 + b_4 f & (y \gg 1).
\label{r-L_theory_smallF}
\end{cases}
\end{eqnarray}
where $b_1=R$, $b_2 =\lambda_p( L_0 +a_1/2a_2 +\ln{(4\pi)}/4a_2\lambda_p)$, $b_3 = a_1/2a_2+L_0+R$ and $b_4 =1/2a_2$.

In the first limiting case ($y \ll 1$), the unbound chain section is coiled in space, where the chain extension is linear as expected (linear response to the weak external field). 
In the freely jointed chain model, the response coefficient of the chain with contour length $L$, and the segment length $l$ is easily calculated from eq. (\ref{r-f_freely-jointed}) as $L l /2$.
However, in the sphere-chain complex, the contour length of the unbound chain section is not fixed but dependent on conditions.
Our theory claims that the response of the disordered sphere-chain complex to the weak stretching force is dictated by entropic elasticity of the unbound chain, whose contour length is determined by the balance between energetic gain due to adsorption (and other terms proportional to $L$) and inter-segment repulsion on the core.
It should be noted that the theory underestimates the chain extension for very low stretching force.
This deviation probably comes from the steric repulsion between the core particle and unbound chain segments, which is not taken into account in our theory.

The second limit ($y \gg 1$) corresponds to the situation, where the unbound chain section is completely stretched.
Such situation should be realized if the attractive interaction between segments and the core particle ($\epsilon$) is sufficiently strong.
Eq. (\ref{r-L_theory_largeF}) indicates that, in this limit, too, the chain extension under large stretching force is a {\it linear} function with the applied tension with a slope $b_4$.
However, the origin of the linearity stems from the different physics from that in the weak stretching force regime.
In this limit of the complete stretching of unbound chain, the chain extension is linearly related to the length of the adsorbed chain section.
In the disordered complex, the length of the adsorbed chain section is determined by the balance between energetic gain due to adsorption (and other terms proportional to $L$) and inter-segment repulsion on the core. 
When the external tension is applied, the system responds by shifting this balance, where the response coefficient is given by $b_4$.

It is also intriguing to consider the situation, in which attractive interactions work between segments.
In this case, the second virial coefficient $B$ of segment-segment interaction is expected to decrease and to change its sign with the increase of the segment-segment attraction.
At that ``$\theta$-point'', the third virial coefficient $C$, instead of $B$, becomes relevant term in eq. (\ref{F_disorder}), and as a consequence, we have a term proportional to $L^3$, instead of $L^2$, in the conditional free energy $G_d(L)$ (eq. (\ref{G_disorder2})).
As a result, a {\it non-linear} response to the stretching force is expected at ``$\theta$-conditions''.

\subsection{Unwrapping transition in an ordered complex}
Contrary to the disordered structure made from a random adsorption of a flexible chain, a stiff chain forms an ordered complex with a core, 
where the segment fluctuations around the optimal configuration are negligible.
In this case, a theoretical argument based on a ground state analysis indicates the all-or-none unwrapping transition\cite{Marky_BioP}. 
Consider a stiff chain wrapping around a cylindrical core particle.
The free energy of this cylinder-wrapped chain complex can be written as a sum of the adsorption and the bending energy terms.
\begin{eqnarray}
F_o (L)= F_{ad}+F_{bend},
\end{eqnarray}
where $F_{ad}=\epsilon  L$ and $F_{bend} = \kappa  L /2 R_0^2$.
$L$ and $R_0$ denote the chain length and radius of curvature, respectively, of the wrapped part.
Both terms are linear in the wrapped chain length $L$.
This indicates that, by changing either the adsorption energy density $\epsilon (\epsilon <0)$ or chain rigidity $\kappa$, one expects the following two situations; the complete wrapping (the chain wraps around the cylindrical core as much as possible until the adsorption space is exhausted), or the complete unwrapping (the chain takes a stretched configuration due to the chain rigidity).
The boundary between these two regimes is given as $\epsilon = -\kappa/2R_0^2$.

In our system, too, the ground state analysis is expected to be a good approximation for the ordered complex.
When the tension is weak, the entropic elasticity of the unbound part play an important role as is discussed in Sec. \ref{sec_disorder}.
In this regime, the chain extension increases rather rapidly with the tension.
We note that this regime of weak stretching force becomes narrower with the increase in the chain stiffness, since the relevant parameter is $y \equiv 2 \lambda_p f$ (see eq. (\ref{r-L_theory_largeF})), which is indeed confirmed by the simulation (Fig. \ref{f-r}).

Here we consider another limiting case, where the unbound chain part is completely stretched by a strong tension.
Under such circumstances, the free energy of the ordered complex with the applied tension $f$ is written as
\begin{eqnarray}
G_o(L) = F_o (L) -f \cdot (L_0-L) .
\label{G_order}
\end{eqnarray}
Since the added term due to the stretching force is also  linear in $L$, the all-or-none unwrapping transition is expected in this case, too. 
However, the following differences should be noted; (i) the presence of stretching force acting on the chain. 
(ii) the possibility of multiple wrapping of the chain around a spherical core.
As for the first point, the stretching force strongly restricts the chain segment around the boundaries between adsorbed and free segments, i.e., entry and exit points.
This results in the discrete unwrapping transition between states with different wrapping numbers separated by an energy barrier, which is discussed below. 
And the second point leads to the step-wise, but not all-or-none (predicted for a cylindrical core), unwrapping transition.
The step-wise unwrapping is a consequence of the spherical geometry of a core particle, in which the bending energy contribution depends on the wrapping number due to the curvature of the core and the hard core volume of the chain.

To elucidate the first point, i.e., the effect of the stretching force on the energy barrier, we study the dynamical process of the unwrapping transition by stretching the end monomer slowly by constant speed $v$.
We adopt $v \Delta t =0.0002$, which is, in fact, very slow compared to the diffusion.

\begin{figure}[h]
\epsfig{figure=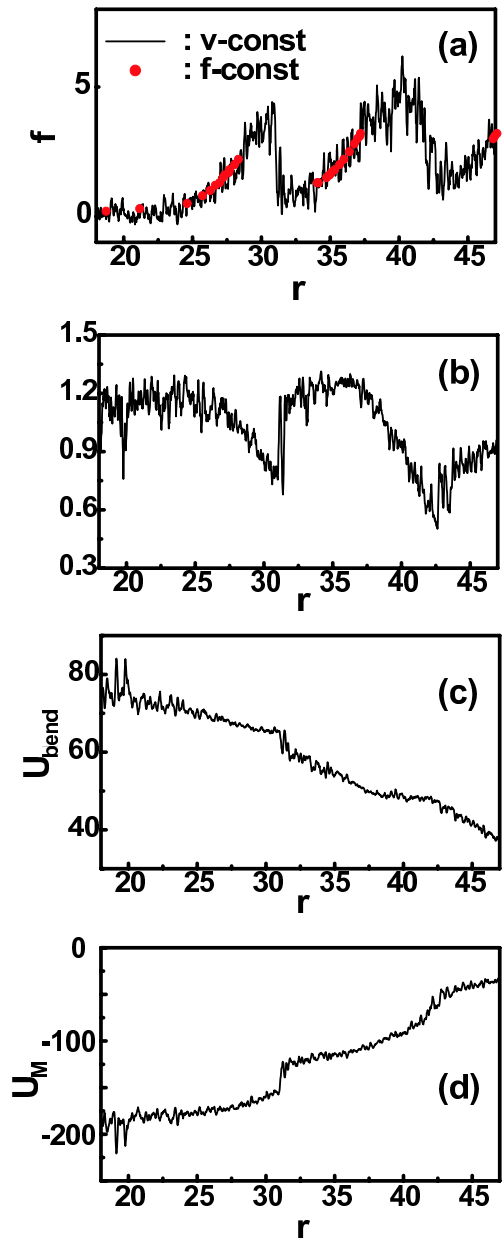,width=6.5cm}
\caption{Unwrapping process by stretching an end monomer by a constant speed for the chain stiffness parameter $\kappa = 10$. (a) Force-extension relation. The result from the force constant ensemble is also plotted by circles for comparison. (b) Tilting angle-extension relation. (c) Bending energy-extension relation. (d) Adsorption energy-extension relation.}
\label{constV}
\end{figure}
Figure \ref{constV} (a) shows the force-extension curve for the case with stiff chain of $\kappa=10$, where the characteristic saw-tooth pattern is observed.
For comparison, the correspondence with the result from the constant force ensemble (Fig. \ref{f-r}) is also included.
This clearly demonstrates the energy barrier during the unwrapping transition, and that quite a large force is required to unwrap the chain.
To see the behavior of the complex during the unwrapping process by the stretching force, we monitor the tilting angle of the ordered complex $\phi$, which is defined by the angle between the axis of the helical pathway of the chain on the sphere and the direction of the force ($x$ axis), as is depicted in Fig. \ref{schematic}.
The helical axis is calculated by a sum of cross products of bond vectors in the vicinity of the core particle's surface (see also the definition of the order parameter $\eta$ given in eq. (\ref{orderparameter})).
Figure \ref{constV} (b) shows the tilting angle and extension relation.
Because of the finite thickness of the chain, the vector connecting the monomers at entry and exit points of the complex is not parallel to the force direction, which produces a torque working on the complex.
Owing to this torque, the complex maintains almost constant tilting angle, which slightly deviates from the perpendicular angle: $\phi \simeq 1.1 \sim 1.2$ up to $r \simeq 27$.
By further stretching the chain beyond $r \simeq 27$, the tilting angle shows a sudden decrease.
This point exactly corresponds to the region where the resistant force sharply increases.
These observations indicate the unique feature of unwrapping process of the ordered sphere-chain complex by the stretching force.
It is not a gradual peeling of the chain around the core, but a chain unfolding accompanied by the complex tilting.
Comparing the Fig. \ref{constV} (a), and Fig. \ref{constV} (b), the complex in the large tilting angle is not in the global minimum of free energy, but in the metastable state, if the stretching force does not exceed the threshold of the limit of stability.
However, this metastable state is blocked by an energy barrier, which should be overcome by a further tilting to reach the globally stable state.
Therefore, even if the stretching force larger than a critical strength for the unwrapping transition is applied, the sphere-chain complex stays in a kinetically stabilized state during a finite life time.

Figure \ref{constV} (c), (d) show the bending energy $U_{bend}$, adsorption energy $U_M$ v.s. extension relations, respectively.
In the region of large tilting, $r\simeq 27 \sim 31$, and $r\simeq 37 \sim 43$, these quantities behave differently from the other regions.
The change in $U_{bend}$ becomes small, or even almost constant, while the change in $U_M$ becomes sharper.
According to these observations, in the tilted state, the number of monomers adsorbed on the core is diminished, thus, energetically less favored, and nevertheless, the cost in the bending energy is not changed, indicating this state is unstable in terms of bending, too\cite{Kulic_cm}.

In the experiments of chromatin stretching by optical tweezers, similar saw-tooth patterns in force-extension curves have been reported\cite{Leuba_PNAS,Cui_PNAS,Bennink_NatureStBio,Brower-Toland_PNAS}.
The measured force peaks are attributed to the unwrapping of the nucleosomes, and the strength of the force peaks was found to be $\simeq 20 pN$, which is almost $10$ times larger than the estimated value from the equilibrium theory\cite{Marko_BioPhysJ,Kunze_PRE,Schiessel_JPCM_review}.
To understand these observations, the inevitability of the nucleosome tilting under the stretching force has been suggested\cite{Cui_PNAS}, and the corresponding energy barrier has also been calculated\cite{Kulic_cm}.
Although this feature is expected to be enhanced in the system with specific site-site interactions as in the case of nucleosome, our results confirm that it is a general one inherent in unwrapping process of an ordered core-chain complex under the tension.

\subsection{Tension induced ordering for intermediate chain stiffness}
In this subsection, we investigate the possibility of tension induced ordering, which is observed in the simulation with an intermediate chain stiffness $\kappa =5$.
It is expected that the ordering by the stretching force is induced by a rather strong force ($y \gg 1$), as is indeed the case in the simulation.
In the regime of strong forces, the free energy of the disordered state under the stretching force $f$ (eq. (\ref{G_disorder2})) is transformed into the following simple form:
\begin{eqnarray}
G_d (L) = (a_1'+f) L + a_2 L^2 + const., 
\end{eqnarray}
where $a_1' \equiv a_1 + \Delta a_1$.
$\Delta a_1$ denotes the bending energy associated with the disordered segment distribution, which should be taken into account, since the chain under consideration has some stiffness.
It is approximated as $ \Delta a_1\sim \kappa  /{R'}^2$.
$R'$ is the effective mean radius of curvature of the chain segments on the core, and should be determined self-consistently for a quantitative discussion.
However, on a qualitative level, it is enough to notice that $\Delta a_1$ is small for a flexible chain and becomes substantially large for a stiff chain.
After representing the system by the most probable state, which is derived by $\partial G_d (L) / \partial L =0$, (mean field approximation), we obtain the optimum free energy for the disordered complex:
\begin{eqnarray}
G_d^* = -\frac{(a_1'-f)^2}{4a_2} + const.
\label{G_disorder_largeF}
\end{eqnarray}
It is noticeable that this is a {\it quadratic} with respect to $f$.

On the other hand, the free energy of the ordered state under stretching force is given by eq. (\ref{G_order}), which is a {\it linear} function of $f$.

The position and the form of these free energies as functions of $f$ depend on the chain stiffness and the size of the core particle.
For a flexible chain and a large core, $G_d^* < G_o$ independent on the value of $f$, while $G_d^* > G_o$ for a chain with sufficient stiffness and a small core.
However, the crossover from the one state to another is expected for the intermediate chain stiffness at a critical force, where $G_d^* = G_o$ holds.
This is a consueqence of the dependencies of free energies on $f$ (eq. (\ref{G_disorder_largeF}), (\ref{G_order})), and indeed corresponds to the tension induced disorder-order transition.

\section{Conclusions}
In conclusion, we have investigated a simple model for a complex formed by a globular protein which attracts a DNA chain. 
Depending on the model chain stiffness, the complex is either disordered on the spherical surface or exhibits an order, i.e.\ a clear winding around the spherical core. 
We have studied by computer simulation and simple phenomenological theory, the unwrapping of this complex under an external stretching force which acts on one end monomer of the chain, while the sphere and the other end monomer are kept fixed.
For a disordered complex, we observe a linear scaling between the force applied and the amount of stretching achieved. 
This linear force-distance law is quantitatively confirmed by a simple 
phenomenological approach in the case where the forces are relatively high. 
On the other hand, for large chain stiffness when there is a clear winding, there are abrupt unwrapping transitions which are connected to a change of the winding number of the chain around the sphere. 
If a constant velocity to one end monomer is applied, there is a large energetic barrier in unspooling the complex. 
Moreover the unwinding is accompanied by a tilt of the spooled complex relative to the pulling direction. 
Finally we have observed the effect of tension-induced ordering for intermediate chain stiffness: If a disordered complex is stretched, the external stretching force induces ordering, i.e.\ winding around the sphere. 

Our model is in principle be realized in chromatin stretching experiments \cite{Leuba_PNAS,Cui_PNAS,Bennink_NatureStBio,Brower-Toland_PNAS} where the protein and the end monomers are guided by optical tweezers. 
We think that pulling experiments of weakly bound complexes could be interpreted in the framework of our disordered situation with a linear force-distance behaviour. Another realization of our model on a larger length scale are complexes of polyelectrolytes and spherical charged colloidal particles \cite{Dzubiella_Macro, Loewen_JPA}, where the persistence length of the chain can be tuned via the salt concentration in the solution \cite{Barrat_Adv.Chem.Phys}.

\section{Acknowledgements}
We thank H. Schiessel and I.M. Kuli{\'c} for helpful comments. This work was supported from the DFG under contract LO 418/9 and a Grant-in-Aid for JSPS fellows from the Ministry of Education, Culture, Sports, Science and Technology of Japan.

\bibliography{sphere-chain}

\newpage




\clearpage




\end{document}